\documentclass[a4paper,fleqn,usenatbib]{mnras}

\usepackage[T1]{fontenc}
\usepackage{ae,aecompl}

\usepackage{graphicx}	
\usepackage{amsmath}	
\usepackage{amssymb}	

\newcommand{\pcc}{\,{\rm cm}^{-3}}
\newcommand{\pcs}{\,{\rm cm}^{-2}}
\newcommand{\um}{\, {\rm \mu m}}
\newcommand{\kel}{\, {\rm K}}
\newcommand{\kk}{\, {\rm kK}}
\newcommand{\lsun}{\, {\rm L}_\odot}
\newcommand{\nh}{n_{\rm H}}
\newcommand{\sfb}{\, {\rm erg} \, {\rm cm}^{-2} \, {\rm s}^{-1} \, {\rm sr}^{-1}}
\newcommand{\flux}{\, {\rm erg} \, {\rm cm}^{-2} \, {\rm s}^{-1}}
\newcommand{\pc}{\, {\rm pc}}

\title[OH$^+$ in PNe]{OH$^+$ emission from cometary knots in planetary nebulae}

\author[F. D. Priestley \& M. J. Barlow]{
F. D. Priestley$^{1}$
and M. J. Barlow$^{1}$
\\
$^{1}$Department of Physics and Astronomy, University College London, Gower Street, London WC1E 6BT, UK\\
}

\date{Accepted XXX. Received YYY; in original form ZZZ}

\pubyear{2018}

\begin{document}
\label{firstpage}
\pagerange{\pageref{firstpage}--\pageref{lastpage}}
\maketitle

\begin{abstract}

  We model the molecular emission from cometary knots in planetary nebulae (PNe) using a combination of photoionization and photodissociation region (PDR) codes, for a range of central star properties and gas densities. Without the inclusion of ionizing extreme ultraviolet (EUV) radiation, our models require central star temperatures $T_*$ to be near the upper limit of the range investigated in order to match observed H$_2$ and OH$^+$ surface brightnesses consistent with observations - with the addition of EUV flux, our models reproduce observed OH$^+$ surface brightnesses for $T_* \ge 100 \kk$. For $T_* < 80 \kk$, the predicted OH$^+$ surface brightness is much lower, consistent with the non-detection of this molecule in PNe with such central star temperatures. Our predicted level of H$_2$ emission is somewhat weaker than commonly observed in PNe, which may be resolved by the inclusion of shock heating or fluorescence due to UV photons. Some of our models also predict ArH$^+$ and HeH$^+$ rotational line emission above detection thresholds, despite neither molecule having been detected in PNe, although the inclusion of photodissociation by EUV photons, which is neglected by our models, would be expected to reduce their detectability.

\end{abstract}




\section{Introduction}

Planetary nebulae (PNe) are the end-stage of the life cycle of low- and intermediate-mass stars, consisting of the gas ejected during previous phases of stellar evolution, surrounding a hot central star. Following the discovery of CO and H$_2$ emission from PNe \citep{mufson1975,treffers1976}, surveys have established that both molecules are commonly present in PNe (e.g. \citealt{huggins1996,hora1999}), with the molecular gas component in some cases forming a significant fraction of the total nebular mass. Other molecules, such as HCN, HCO$^+$ and CS, have also been detected in PNe \citep{edwards2014,schmidt2016}, allowing investigation of the otherwise undetectable neutral regions.

In order to survive the ultraviolet (UV) radiation from the central star, molecules must be shielded within regions with significant optical depths \citep{tielens1993,natta1998}, such as the cometary knots commonly observed in PNe \citep{odell1996,odell2002}. \citet{matsuura2009} found that the H$_2$ emission from NGC 7293 is highly localised within these knots, although in NGC 6781 the molecular emission more resembles a ring surrounding the central star \citep{bachiller1993,hiriart2005}. \citet{aleman2011} were able to reproduce the observed H$_2$ surface brightnesses of cometary knots in NGC 7293 with a combined photoionization/photodissociation region (PDR) code, finding that the low-density/diffuse gas surrounding the knots contributes very little to the overall H$_2$ emission. A detailed model of NGC 6781, assuming a dense, shock-heated PDR surrounding the inner ionized region, was produced by \citet{otsuka2017} to fit a wide range of observational data from the UV to radio, including H$_2$ and CO molecular emission.

The molecular ion OH$^+$, originally observed in emission around ultraluminous galaxies \citep{vanderwerf2010}, in the ISM \citep{wyrowski2010}, the Orion bar \citep{vandertak2013} and, along with ArH$^+$, in the Crab Nebula \citep{barlow2013}, was detected in five PNe by \citet{etxaluze2014} and \citet{aleman2014}. \citet{aleman2014} noted that all five PNe had central star temperatures greater than $100 \kk$, and suggested that soft X-rays may be responsible for producing this emission, as in the case of ultraluminous galaxies \citep{vanderwerf2010}. X-rays were also found to be needed to reproduce the observed abundances of CN and HCO$^+$ \citep{kimura2012} in PNe. In their model of NGC 6781, \citet{otsuka2017} predicted a significantly larger quantity of OH$^+$ than deduced from observations, despite good agreement with other molecular abundances.

In this paper, we model the molecular emission from cometary knots in PNe using a combination of photoionization and PDR codes. Previous efforts \citep{aleman2004,aleman2011,kimura2012,otsuka2017} have treated both problems simultaneously, ensuring that the PDRs are modelled self-consistently by solving the full radiative transfer problem throughout the knot. We use the approach adopted by \citet{priestley2017} for their Crab Nebula modelling, assuming that the photoionized region can be modelled separately from the PDR, providing an incident radiation field which determines the properties of the neutral gas. This allows us to use more detailed PDR models, including more molecular species and calculating the emission self-consistently, and to investigate a larger range of parameter space, due to the lower computational cost. Rather than attempt to reproduce the properties of one specific object, we vary the input parameters of our models and compare the resulting molecular emission predictions to the values typically observed in PNe.

\section{Method}

To determine the incident radiation field on the knots, we model the photoionized region of the PNe using {\sc mocassin} \citep{ercolano2003,ercolano2005,ercolano2008}, a Monte Carlo photoionization code. We assume a 1D spherically symmetric geometry, with an inner region of diffuse gas of density $n_d$ extending up to a radius $r_k$ at which the knot is located, beyond which the gas density is increased to $n_k$. We increase the width of this dense region until the ionizing extreme UV (EUV) flux (taken to be $200-912$~\r{A}) has been fully absorbed, marking the transition from a photoionized region to a PDR. We then integrate the emerging spectral energy distribution (SED) from $912$ to $2400$~\r{A} to determine the far UV (FUV) field (in Draine units, $\sim 1.7$ times the Habing field; \citealt{habing1968,draine1978}) and from $0.1$ to $200$~\r{A} to determine the X-ray flux incident on the PDR. We then model the PDR as a 1D slab with constant density $n_k$ and a thickness of $\Delta r_k$ using {\sc ucl\_pdr} \citep{bell2005,bell2006,bayet2011,priestley2017}, a PDR code that is also capable of treating X-rays. We assume $\Delta r_k$ is small enough compared to $r_k$ that we can ignore the effect of geometrical dilution on the radiation field inside the PDR. We take $n_d = 100 \pcc$, at the lower end of the typical density range of $10^2-10^4 \pcc$ for diffuse gas in PNe \citep{osterbrock2006}, and $n_k = 10^5 \pcc$, appropriate for the cometary knots in NGC 7293 \citep{meaburn1998,matsuura2007}. We use $r_k = 0.2 \pc$ and $\Delta r_k = 0.003 \pc$, again consistent with values for the cometary knots in NGC 7293 \citep{matsuura2009}.

The gas-phase elemental abundances used are listed in Table \ref{tab:abun}. We used values typical of PNe \citep{kingsburgh1994,pottasch2005}, with the exception of Mg, Si and Fe, where we assumed solar abundances \citep{lodders2010} depleted by the maximum ISM values from \citet{jenkins2009}. We assumed a dust-to-gas mass ratio of $0.005$, typical of values found for PNe \citep{aleman2011,ueta2014,otsuka2017}. We used a dust composition of half amorphous carbon, half silicate grains and a power law grain size distribution with an exponent of $-3.5$, and minimum/maximum grain radii of $0.005$/$0.25 \um$. The choice of grain composition has been found to have only a small effect on H$_2$ abundances and emission - while the dust-to-gas mass ratio and dust grain size can cause significant variations \citep{aleman2004,aleman2011a,aleman2011}, we do not investigate varying the dust properties in our models.

We investigate central star luminosities of $L_* = 100$ and $1000 \lsun$, and temperatures of $T_* = 50$, $80$, $100$, $120$ and $150 \kk$, covering much of the parameter space relevant to old PNe \citep{vassiliadis1994,millerbertolami2016}. We used model stellar spectra from the TheoSSA database \citep{rauch2003}, using $\log g = 6.0$, $7.0$ and $8.0$ as typical values for PNe without stellar winds or outflows \citep{guerrero2013}, and take $\log g = 7.0$ as our standard value. We found that the transmitted X-ray SED output from our {\sc mocassin} models could be well fitted by a four-component power law. The transmitted UV and X-ray fluxes for each combination of $L_*$, $T_*$ and $\log g$ are listed in Table \ref{tab:ionprop}.

For our PDR modelling we used the chemical network from \citet{priestley2017}, including ArH$^+$ and HeH$^+$. We use the dissociative recombination rate for ArH$^+$ with electrons calculated by Abdoulanziz et al. (in prep.). We included the collisional excitation of OH$^+$ by H using the rate coefficients calculated by \citet{lique2016}, which may become important in lower ionization regions, in addition to collisional excitation by electrons \citep{vandertak2013}. We assume a typical ISM value for the cosmic ray ionization rate ($1.3 \times 10^{-17} \, {\rm s}^{-1}$) and an $A_V$/$N_{\rm H}$ conversion factor of $3.0 \times 10^{-22} \, {\rm mag} \, {\rm cm}^2$, reduced from ISM values by a factor of two to account for our lower dust-to-gas mass ratio. Values of the cosmic ray ionization rate derived from observations can be higher than our adopted value by over an order of magnitude (e.g. \citealt{neufeld2017}), which may become important if the ionization rate exceeds that from the X-ray flux.

We also investigate the likely effects of the EUV ($200-912$~\r{A}) flux, suggested to be responsible for powering H$_2$ emission in NGC 7293 by \citet{odell2007}, by running {\sc mocassin} models consisting of only diffuse gas up to $r_k$, without the denser shell absorbing the ionizing radiation. The transmitted fluxes are listed in Table \ref{tab:ionpropfull}. The `X-ray' fluxes, now defined as over the wavelength range $0.1-750$~\r{A}, are significantly increased for all models, and are much less affected by $T_*$, as the expanded range includes the SED peak for all central star temperatures. The UV fields are slightly reduced from the previous values, as the diffuse UV emission from the dense ionized gas (calculated in {\sc mocassin} but not {\sc ucl\_pdr}) is no longer included. {\sc ucl\_pdr} treats X-rays following the methods described in \citet{meijerink2005} - the extent to which this approach is valid for EUV photons, and the likely impact on our results, is discussed in Section \ref{sec:euv}.

\begin{table}
  \centering
  \caption{Gas-phase elemental abundances, relative to hydrogen, adopted for the PN modelling.}
  \begin{tabular}{cccc}
    \hline
    Element & Abundance & Element & Abundance \\
    \hline
    H & $1.00$ & Mg & $1.5 \times 10^{-6}$ \\
    He & $0.10$ & Si & $1.5 \times 10^{-6}$ \\
    C & $2.5 \times 10^{-4}$ & S & $1.5 \times 10^{-5}$ \\
    N & $1.0 \times 10^{-4}$ & Cl & $1.8 \times 10^{-7}$ \\
    O & $5.0 \times 10^{-4}$ & Ar & $2.0 \times 10^{-6}$ \\
    Ne & $1.0 \times 10^{-4}$ & Fe & $1.5 \times 10^{-7}$ \\
    \hline
  \end{tabular}
  \label{tab:abun}
\end{table}

\begin{table*}
  \centering
  \caption{Central star luminosities $L_*$ and effective temperatures $T_*$, and output UV and X-ray fluxes from our {\sc mocassin} modelling of PNe. The X-ray component includes all transmitted flux between $0.1$~\r{A} and $200$~\r{A}..}
  \begin{tabular}{ccccccccccc}
    \hline
     & & \multicolumn{2}{c}{$\log g = 7.0$} & \multicolumn{2}{c}{$\log g = 6.0$} & \multicolumn{2}{c}{$\log g = 8.0$} \\
    $L_*$/$L_\odot$ & $T_*$/kK & $F_{\rm UV}$/Draines & $F_{\rm X}$/$\flux$ & $F_{\rm UV}$/Draines & $F_{\rm X}$/$\flux$ & $F_{\rm UV}$/Draines & $F_{\rm X}$/$\flux$ \\
    \hline
    $10^2$ & $50$ & $233$ & $0.00$ & $226$ & $0.00$ & $238$ & $0.00$ \\
    $10^2$ & $80$ & $91.5$ & $3.24 \times 10^{-7}$ & $92.9$ & $3.67 \times 10^{-7}$ & $90.6$ & $3.73 \times 10^{-7}$ \\
    $10^2$ & $100$ & $69.8$ & $2.18 \times 10^{-5}$ & $71.5$ & $7.31 \times 10^{-5}$ & $68.7$ & $4.96 \times 10^{-6}$ \\
    $10^2$ & $120$ & $59.2$ & $6.56 \times 10^{-4}$ & $60.0$ & $2.32 \times 10^{-3}$ & $58.8$ & $4.04 \times 10^{-4}$ \\
    $10^2$ & $150$ & $48.1$ & $4.55 \times 10^{-2}$ & $47.2$ & $6.16 \times 10^{-2}$ & $48.7$ & $3.52 \times 10^{-2}$ \\
    $10^3$ & $50$ & $2230$ & $0.00$ & $2160$ & $0.00$ & $2280$ & $0.00$ \\
    $10^3$ & $80$ & $817$ & $1.05 \times 10^{-6}$ & $829$ & $1.01 \times 10^{-6}$ & $815$ & $2.13 \times 10^{-6}$ \\
    $10^3$ & $100$ & $606$ & $2.39 \times 10^{-4}$ & $622$ & $8.61 \times 10^{-4}$ & $596$ & $5.75 \times 10^{-5}$ \\
    $10^3$ & $120$ & $501$ & $7.38 \times 10^{-3}$ & $499$ & $4.19 \times 10^{-2}$ & $497$ & $4.43 \times 10^{-3}$ \\
    $10^3$ & $150$ & $366$ & $0.570$ & $351$ & $0.713$ & $377$ & $0.459$ \\
    \hline
  \end{tabular}
  \label{tab:ionprop}
\end{table*}

\begin{table}
  \centering
  \caption{Central star luminosities $L_*$ and effective temperatures $T_*$, and output UV and X-ray fluxes from our {\sc mocassin} modelling of PNe, for model atmospheres with $\log g = 7.0$ including the EUV flux. The X-ray component now includes all flux up to $750$~\r{A}.}
  \begin{tabular}{ccccc}
    \hline
    $L_*$ / $L_\odot$ & $T_*$ / kK & $F_{\rm UV}$ / Draines & $F_{\rm X}$ / $\flux$ \\
    \hline
    $10^2$ & $50$ & $233$ & $0.183$ \\
    $10^2$ & $80$ & $85.1$ & $0.550$ \\
    $10^2$ & $100$ & $62.7$ & $0.605$ \\
    $10^2$ & $120$ & $51.7$ & $0.613$ \\
    $10^2$ & $150$ & $41.7$ & $0.580$ \\
    $10^3$ & $50$ & $2110$ & $6.54$ \\
    $10^3$ & $80$ & $609$ & $11.8$ \\ 
    $10^3$ & $100$ & $387$ & $12.6$ \\
    $10^3$ & $120$ & $284$ & $12.4$ \\
    $10^3$ & $150$ & $175$ & $13.4$ \\
    \hline
  \end{tabular}
  \label{tab:ionpropfull}
\end{table}

\section{Results}

\begin{figure}
  \centering
  \includegraphics[width=\columnwidth]{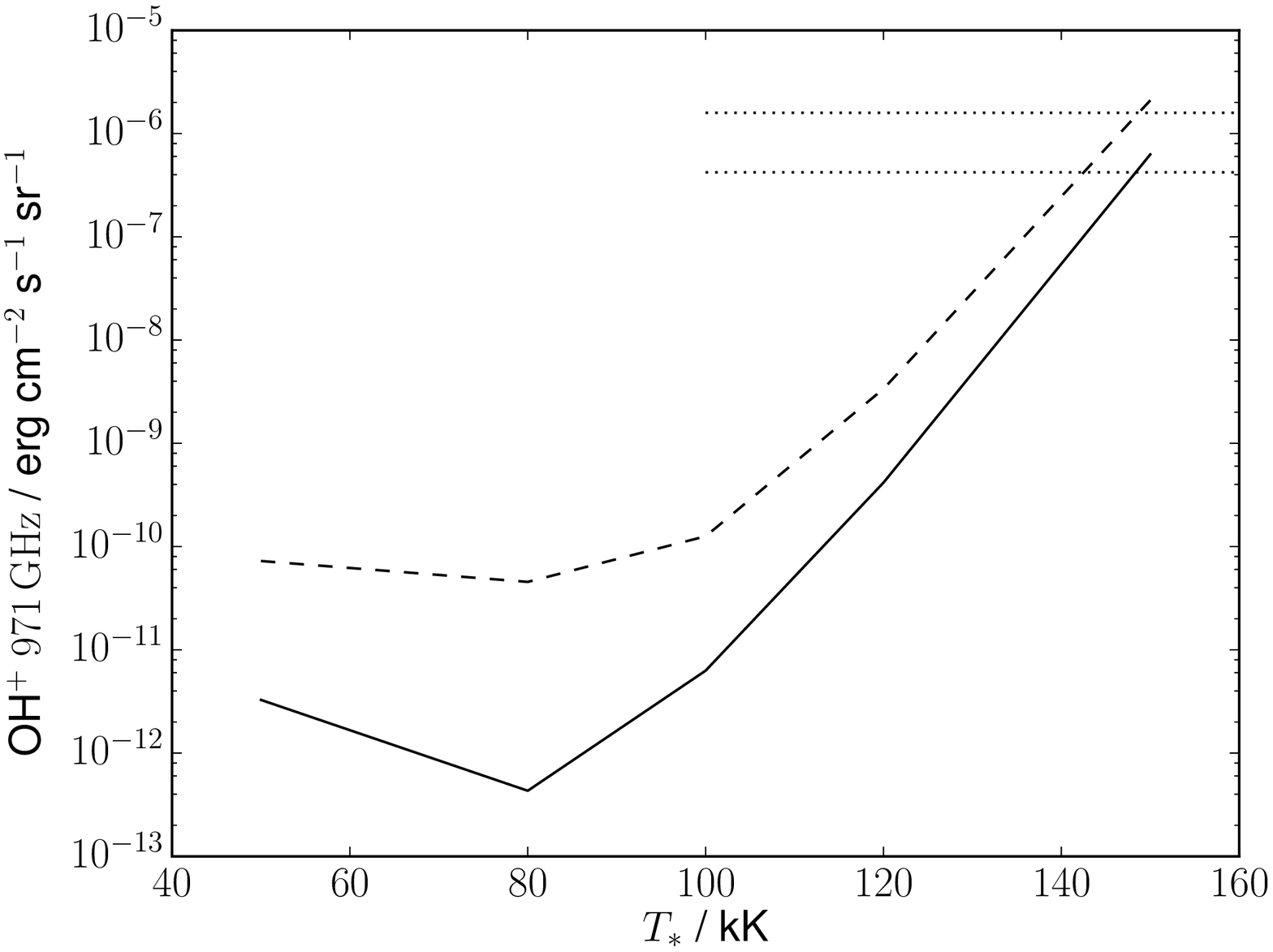}
  \caption{OH$^+$ 971 GHz surface brightness versus $T_*$, for models with $\nh = 10^5 \pcc$, $\log g = 7.0$ and $L_* = 100 \lsun$ (solid line) and $1000 \lsun$ (dashed line). The dotted horizontal lines show the range of observed values from \citet{etxaluze2014} and \citet{aleman2014}.}
  \label{fig:971emis}
\end{figure}

Figure \ref{fig:971emis} shows the predicted OH$^+$ 971 GHz line surface brightness versus $T_*$, for models with $\log g = 7.0$. The emission strength increases rapidly for central star temperatures above $100 \kk$, while also increasing with $L_*$, as increasing X-ray fluxes enhance the ionized fraction in the gas and thus the OH$^+$ abundance \citep{vanderwerf2010}. The observed range of surface brightnesses \citep{aleman2014,etxaluze2014} can be explained by models with central star temperatures close to $150 \kk$, although these models have OH$^+$ column densities in excess of the values calculated by \citet{aleman2014}. For $T_* \le 120 \kk$ the surface brightnesses are well below those observed, despite several PNe with OH$^+$ detections having central star temperatures in this region. The line ratios between the three lowest frequency OH$^+$ rotational transitions, at 909, 971 and 1033 GHz, are essentially constant across all models, with relative strengths of $0.2:1:0.6$ respectively. This is in qualitative agreement with the ratios for all the PNe with detected OH$^+$ emission, from \citet{aleman2014} and \citet{etxaluze2014}, with the exception of NGC 6781.

\begin{figure}
  \centering
  \includegraphics[width=\columnwidth]{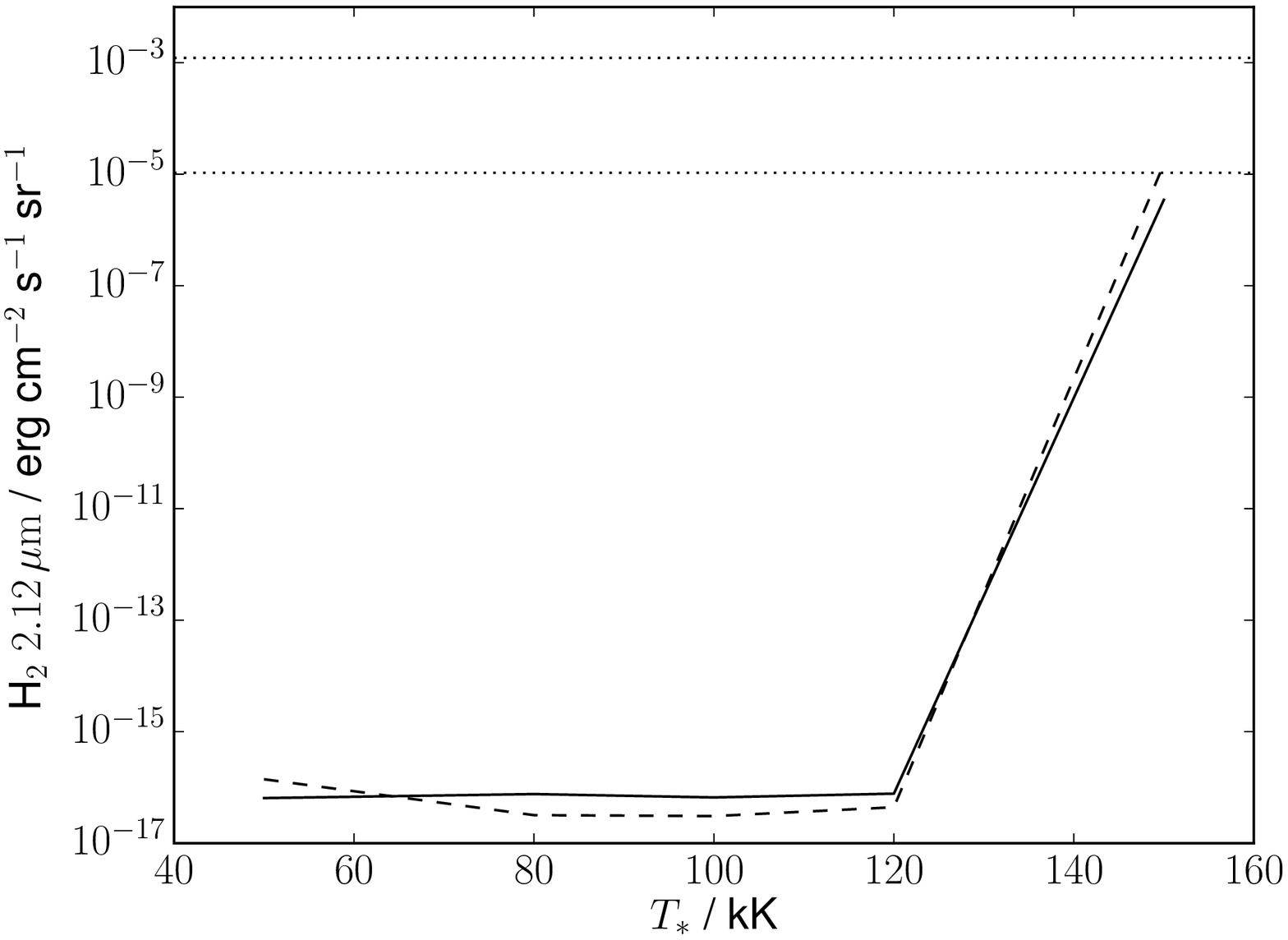}
  \caption{H$_2$ $2.12 \um$ surface brightness versus $T_*$, for models with $\nh = 10^5 \pcc$, $\log g = 7.0$ and $L_* = 100 \lsun$ (solid line) and $1000 \lsun$ (dashed line). The dotted horizontal lines show the range of observed values from \citet{hora1999}.}
  \label{fig:212emis}
\end{figure}

Figure \ref{fig:212emis} shows predicted the H$_2$ $2.12 \um$ line surface brightness versus $T_*$ for two values of the central star luminosity. The range of values observed in PNe from \citet{hora1999} is also shown. The predicted emission strength is barely affected by $L_*$, and is roughly constant below $120 \kk$, as the H$_2$ line requires significant population of the $v = 1$ vibrational state, and the PDR gas temperature is far higher for the $150 \kk$ models than for those with lower $T_*$. For $T_* = 150 \kk$, the predicted H$_2$ surface brightnesses are consistent with the smallest values detected by \citet{hora1999}, but the highest surface brightnesses in this line are not reproduced by these models.

\begin{figure}
  \centering
  \includegraphics[width=\columnwidth]{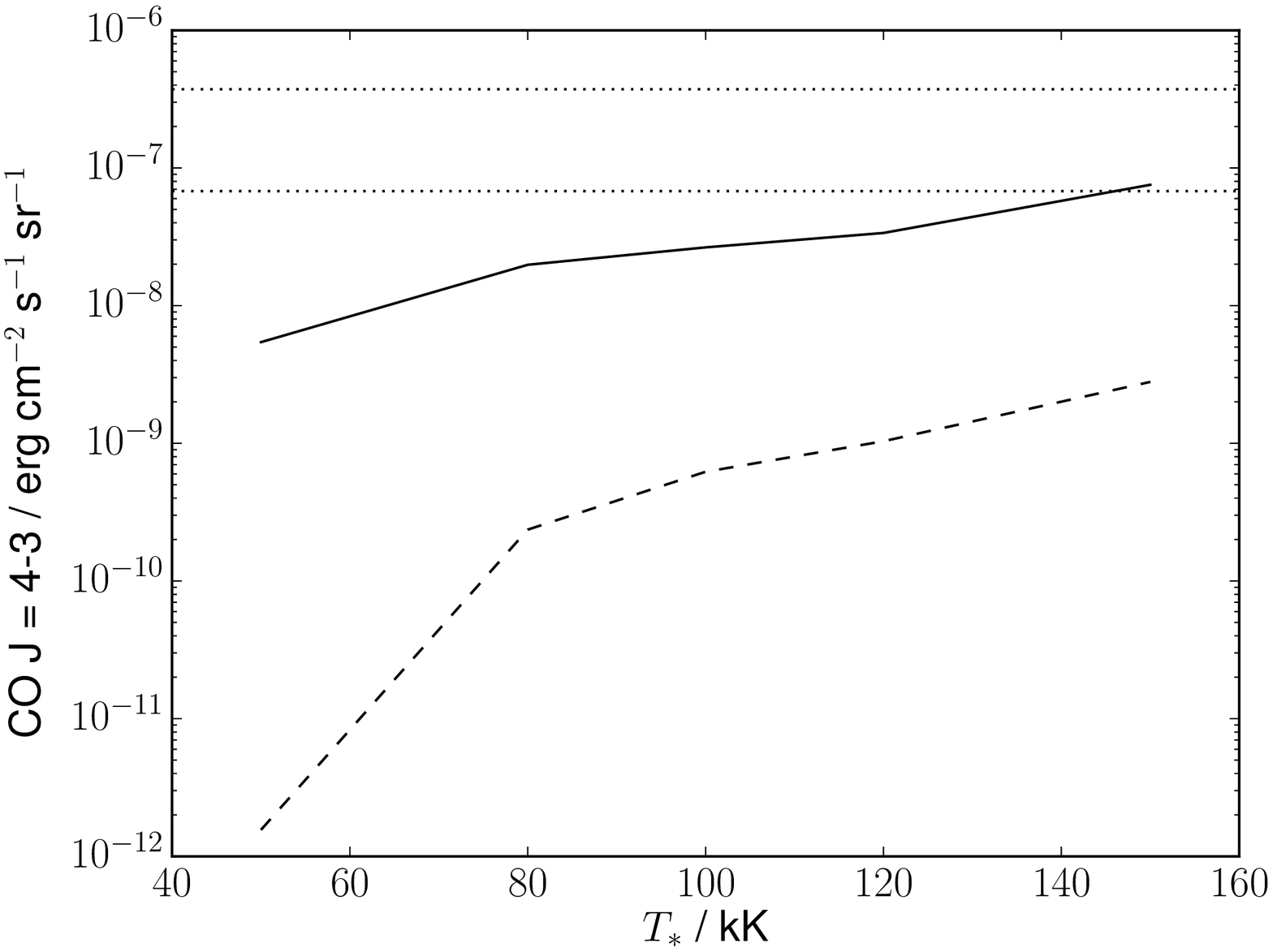}
  \caption{CO $J=4-3$ surface brightness versus $T_*$, for models with $\nh = 10^5 \pcc$, $\log g = 7.0$ and $L_* = 100 \lsun$ (solid line) and $1000 \lsun$ (dashed line). The dotted horizontal lines show the range of observed values from \citet{etxaluze2014} and \citet{ueta2014}.}
  \label{fig:co43emis}
\end{figure}

Figure \ref{fig:co43emis} shows the predicted CO $J=4-3$ line surface brightness versus $T_*$. In this case the emission is strongly affected by the value of $L_*$, as the increased UV field for the $1000 \lsun$ models readily dissociates CO molecules. Surface brightness increases with $T_*$ for the same $L_*$ as less of the central star flux is emitted at FUV wavelengths. The $100 \lsun$, $150 \kk$ model has a predicted surface brightness comparable with the lowest values measured by \citet{etxaluze2014} and \cite{ueta2014}, but again, no model reproduces the higher observed values. Increasing the elemental abundance of carbon is unlikely to resolve this discrepancy, as our oxygen abundance is only a factor of two higher, limiting the amount of CO which can be formed. Elemental oxygen abundances in PNe are rarely significantly higher than our value of $5 \times 10^{-4}$ \citep{kingsburgh1994}.

\begin{figure}
  \centering
  \includegraphics[width=\columnwidth]{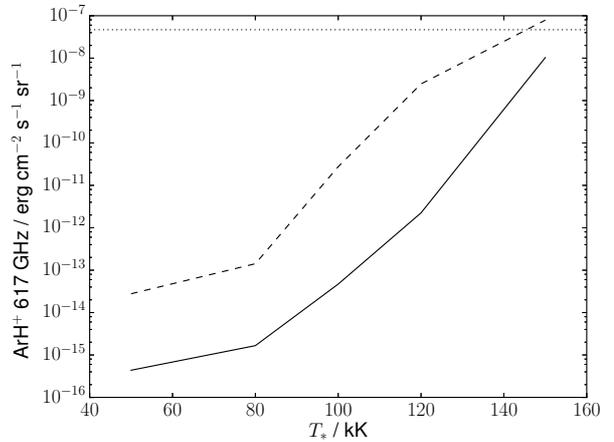}
  \caption{ArH$^+$ 617 GHz surface brightness versus $T_*$, for models with $\nh = 10^5 \pcc$, $\log g = 7.0$ and $L_* = 100 \lsun$ (solid line) and $1000 \lsun$ (dashed line). The dotted horizontal line shows the surface brightness of the weakest SPIRE OH$^+$ detection from \citet{aleman2014}.}
  \label{fig:617emis}
\end{figure}

Figure \ref{fig:617emis} shows the predicted ArH$^+$ $J=1-0$ 617 GHz line surface brightness versus $T_*$. ArH$^+$ has not been detected in any PN to date, despite the $J=1-0$ and $2-1$ transitions falling within the Herschel SPIRE frequency range. The dashed line in Figure \ref{fig:617emis} corresponds to the surface brightness of the weakest OH$^+$ SPIRE line detection from \citet{aleman2014}. If this is taken as a lower limit, only the $150 \kk$, $1000 \lsun$ model predicts detectable ArH$^+$ emission in this line, although the $150 \kk$, $100 \lsun$ model is only a factor of a few below the limit. The $J=2-1$ transition at 1234 GHz is generally brighter in all models by a factor of a few, but considering the lower signal-to-noise at higher frequencies, we still find only the $150 \kk$, $1000 \lsun$ model produces a surface brightness above the weakest OH$^+$ (1033 GHz) line detection. A similar situation is found for the HeH$^+$ $149 \um$ line, using the OH$^+$ $152 \um$ PACS detections as a lower limit for detectability - all models except for $150 \kk$, $1000 \lsun$ are below the detection threshold.

\begin{figure}
  \centering
  \includegraphics[width=\columnwidth]{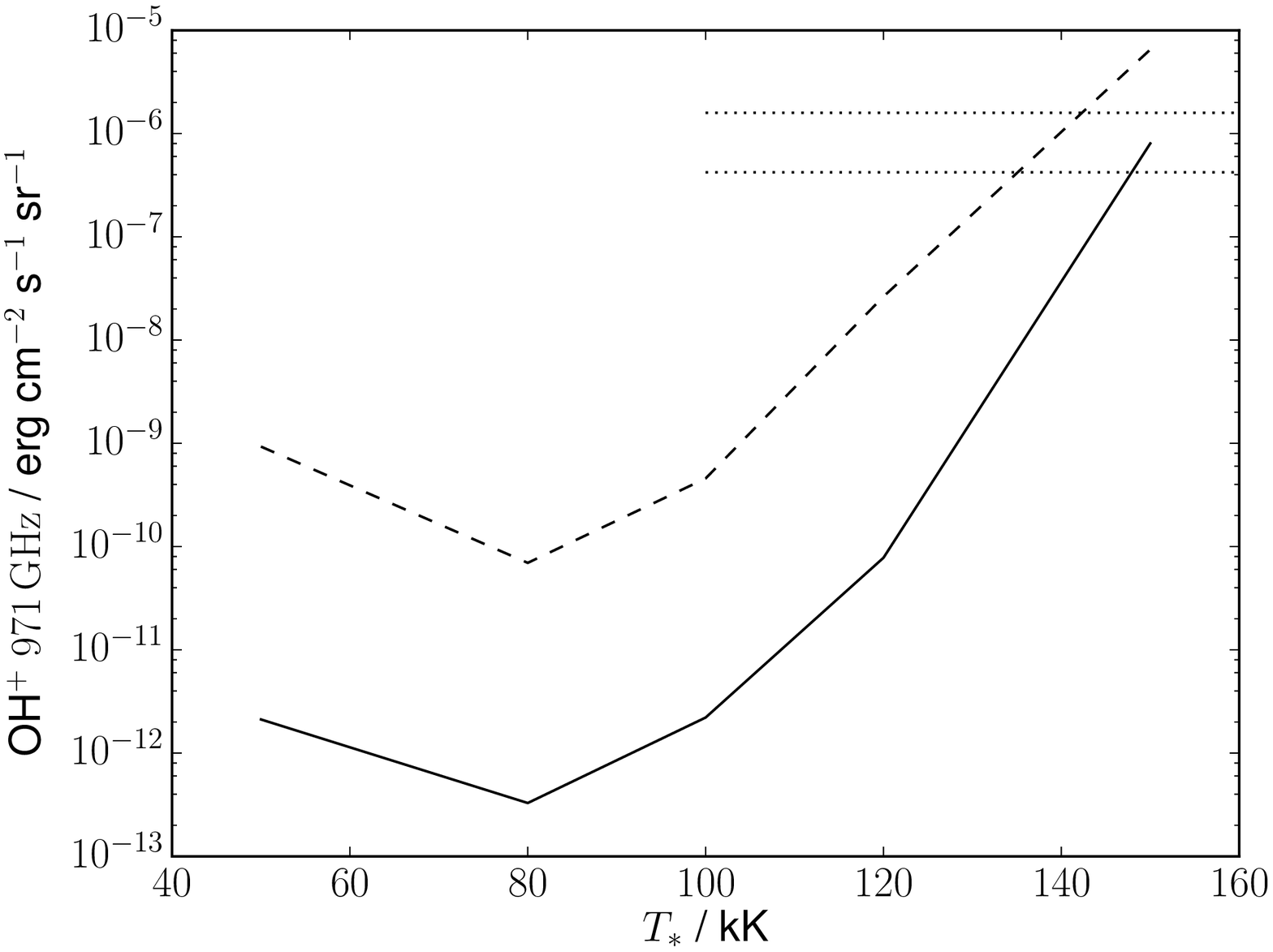}
  \caption{OH$^+$ 971 GHz surface brightness versus $T_*$, for models with $\nh = 10^6 \pcc$, $\log g = 7.0$ and $L_* = 100 \lsun$ (solid line) and $1000 \lsun$ (dashed line). The dotted horizontal lines show the range of observed values from \citet{etxaluze2014} and \citet{aleman2014}.}
  \label{fig:d6971emis}
\end{figure}

\begin{figure}
  \centering
  \includegraphics[width=\columnwidth]{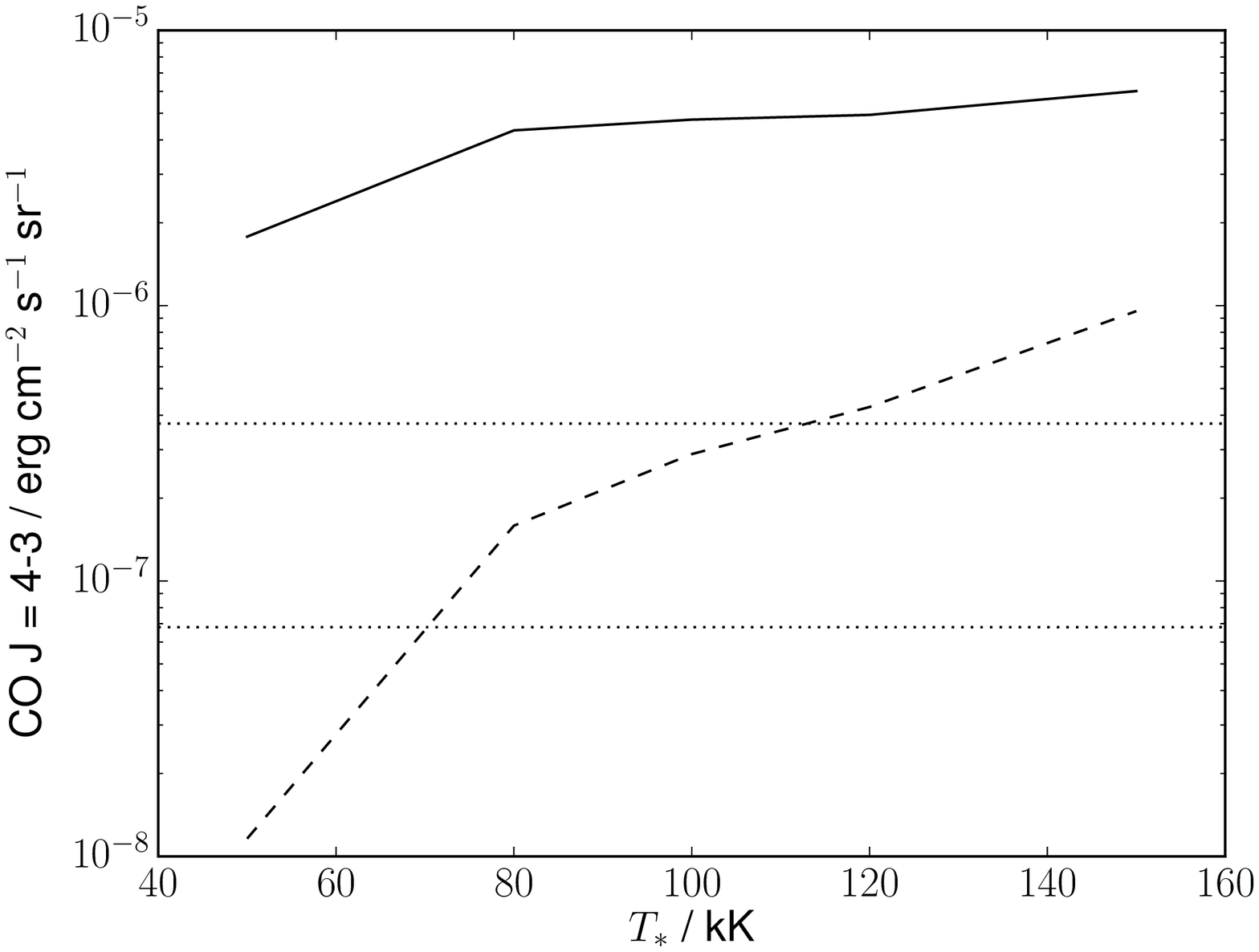}
  \caption{CO $J=4-3$ surface brightness versus $T_*$, for models with $\nh = 10^6 \pcc$, $\log g = 7.0$ and $L_* = 100 \lsun$ (solid line) and $1000 \lsun$ (dashed line). The dotted horizontal lines show the range of observed values from \citet{etxaluze2014} and \citet{ueta2014}.}
  \label{fig:d6co43emis}
\end{figure}

Increasing the gas density to $\nh = 10^6 \pcc$, while reducing $\Delta r_k$ to $3 \times 10^{-4} \pc$ to ensure the same column density, the OH$^+$ and H$_2$ line surface brightnesses are increased for all models, with the effect greater for $L_* = 1000 \lsun$. Figure \ref{fig:d6971emis} shows the OH$^+$ 971 GHz line surface brightness versus $T_*$, for these increased density models. Even for $1000 \lsun$, the increase in surface brightness is not great enough to bring models with $T_* < 150 \kk$ within the range of observed values. The CO $J=4-3$ emission, shown in Figure \ref{fig:d6co43emis}, is much more strongly affected, with all the $100 \lsun$ models now exceeding the observational upper limits, while for $L_* = 1000 \lsun$ the predicted surface brightnesses are comparable to observation, with the exception of the $50 \kk$ model, for which it is still too low. ArH$^+$ and HeH$^+$ surface brightnesses both decrease, although the $150 \kk$, $1000 \lsun$ model still predicts detectable levels for both molecules.

\begin{figure}
  \centering
  \includegraphics[width=\columnwidth]{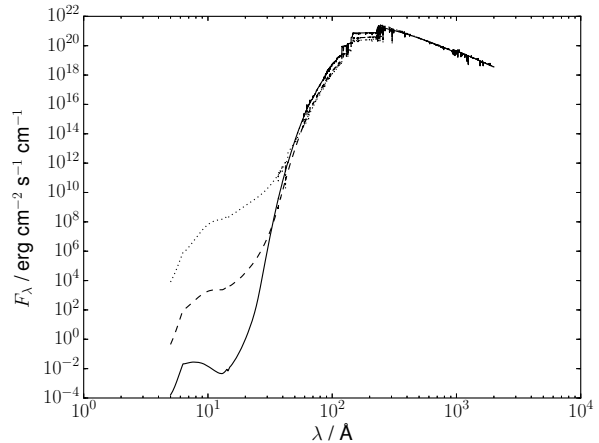}
  \caption{Model central star spectra for model atmospheres with $T_* = 120 \kk$ and $\log g = 6.0$ (solid line), $7.0$ (dashed line) and $8.0$ (dotted line), obtained from the TheoSSA database \citep{rauch2003}.}
  \label{fig:logg}
\end{figure}

Figure \ref{fig:logg} shows the input stellar spectra for model atmospheres with $T_* = 120 \kk$ and $\log g$ ranging from $6.0$ to $8.0$, demonstrating the large variation at X-ray wavelengths between the models (note that the total flux up to $200$~\r{A} is highest for $\log g = 6.0$, despite $F_\lambda$ being much lower at the shortest wavelengths). The results from our photoionization modelling, listed in Table \ref{tab:ionprop}, show that the incident X-ray flux is generally higher for $\log g = 6.0$ and lower for $\log g = 8.0$, while the UV field is not significantly affected.

\begin{figure}
  \centering
  \includegraphics[width=\columnwidth]{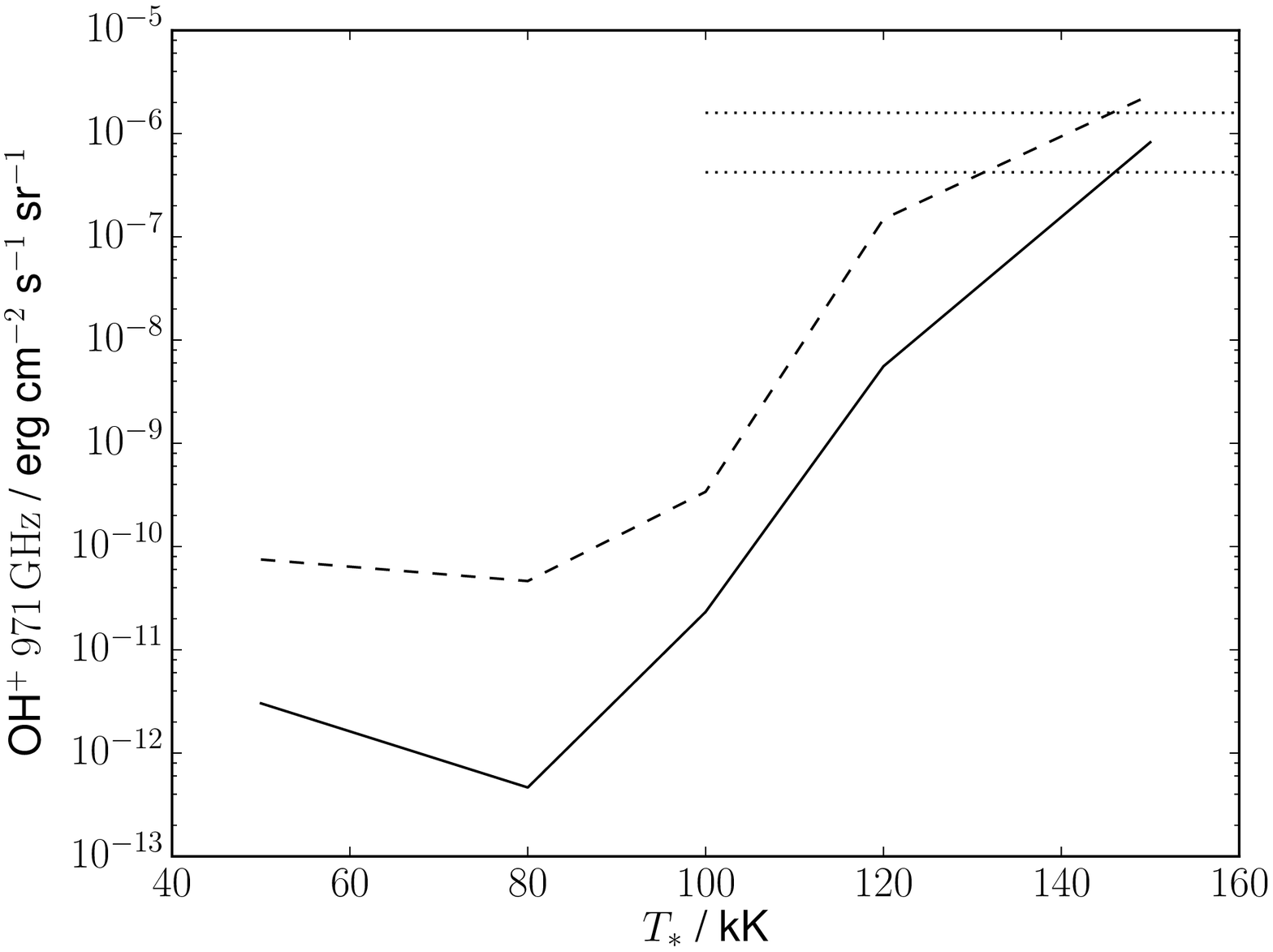}
  \caption{OH$^+$ 971 GHz surface brightness versus $T_*$, for models with $\nh = 10^5 \pcc$, $\log g = 6.0$ and $L_* = 100 \lsun$ (solid line) and $1000 \lsun$ (dashed line). The dotted horizontal lines show the range of observed values from \citet{etxaluze2014} and \citet{aleman2014}.}
  \label{fig:g6971emis}
\end{figure}

Figure \ref{fig:g6971emis} shows the OH$^+$ 971 GHz line surface brightness versus $T_*$, for models with $\log g = 6.0$. The increased X-ray flux results in stronger OH$^+$ emission, particularly for $T_* = 120 \kk$, for which the $1000 \lsun$ model gives a surface brightness within a factor of a few of observations. This model also predicts H$_2$ surface brightnesses consistent with observed values, compared to the $\log g = 7.0$ case, where the emission is far weaker. The CO $J=4-3$ surface brightnesses are largely unchanged from the $\log g = 7.0$ case, while the ArH$^+$ and HeH$^+$ lines are only slightly weaker than the detection limits taken from \citet{aleman2014}. Increasing $\log g$ to $8.0$ reduces model surface brightnesses by at most a factor of a few compared to $\log g = 7.0$.

\begin{figure}
  \centering
  \includegraphics[width=\columnwidth]{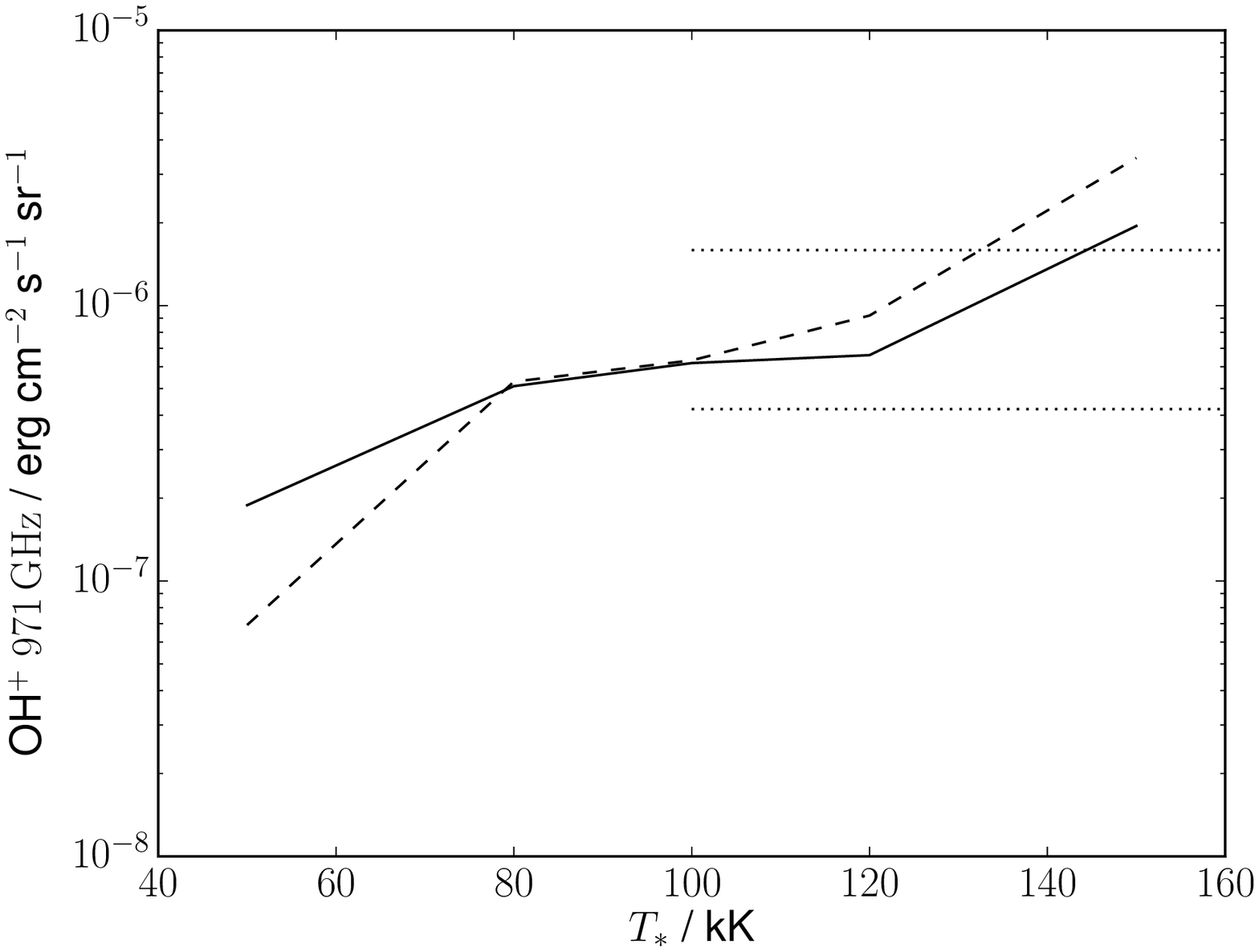}
  \caption{OH$^+$ 971 GHz surface brightness versus $T_*$, for models with $\nh = 10^5 \pcc$, $\log g = 7.0$ and $L_* = 100 \lsun$ (solid line) and $1000 \lsun$ (dashed line) including the EUV flux. The dotted horizontal lines show the range of observed values from \citet{etxaluze2014} and \citet{aleman2014}.}
  \label{fig:full971emis}
\end{figure}

\begin{figure}
  \centering
  \includegraphics[width=\columnwidth]{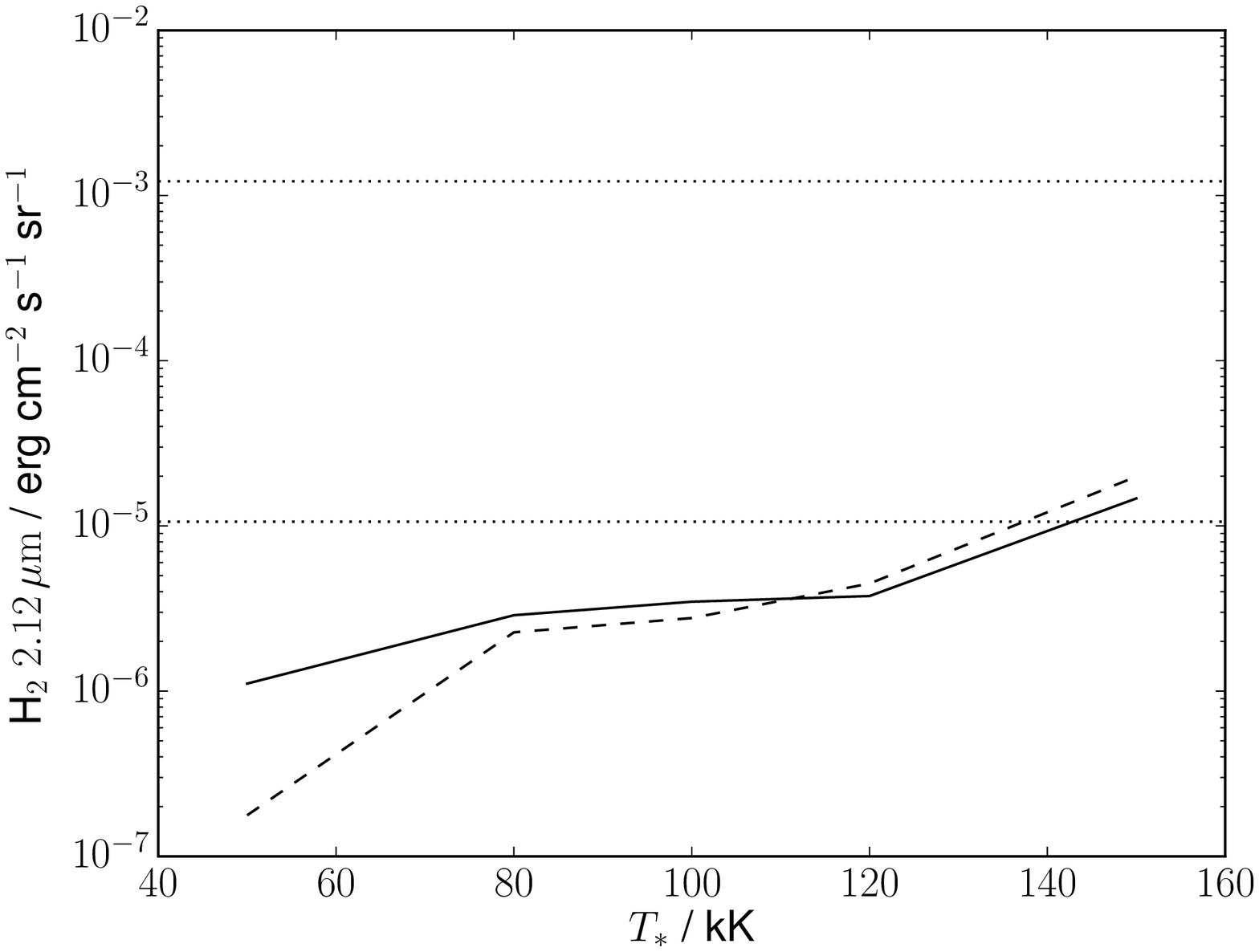}
  \caption{H$_2$ $2.12 \um$ surface brightness versus $T_*$, for models with $\nh = 10^5 \pcc$, $\log g = 7.0$ and $L_* = 100 \lsun$ (solid line) and $1000 \lsun$ (dashed line) including the EUV flux. The dotted horizontal lines show the range of observed values from \citet{hora1999}.}
  \label{fig:full212emis}
\end{figure}

\begin{figure}
  \centering
  \includegraphics[width=\columnwidth]{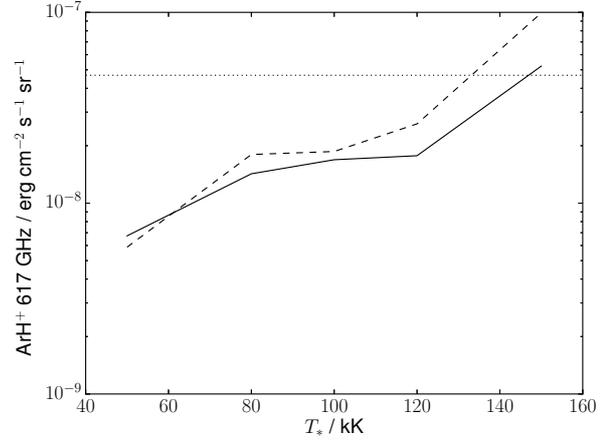}
  \caption{ArH$^+$ 617 GHz surface brightness versus $T_*$, for models with $\nh = 10^5 \pcc$, $\log g = 7.0$ and $L_* = 100 \lsun$ (solid line) and $1000 \lsun$ (dashed line) including the EUV flux. The dotted horizontal line shows the surface brightness of the weakest SPIRE OH$^+$ detection from \citet{aleman2014}.}
  \label{fig:full617emis}
\end{figure}

Figures \ref{fig:full971emis} and \ref{fig:full212emis} show respectively the predicted OH$^+$ 971 GHz and H$_2$ $2.12 \um$ line surface brightnesses versus $T_*$, for models including the EUV flux. Models with $T_* < 150 \kk$ have much higher surface brightnesses in both lines than the corresponding models without the additional flux, as the EUV flux heats the gas at the edge of the knot to much higher temperatures, as well as increasing the ionization fraction. The $H_2$ $2.12 \um$ surface brightnesses are increased to within a factor of a few of observed values for $T_* \ge 80 \kk$. As with H$_2$, for $T_* < 150 \kk$ there is a significant increase in the OH$^+$ surface brightnesses. Models with $T_* > 80 \kk$ all predict surface brightnesses similar to or higher than observations, while for $T_* = 50 \kk$ the predicted values are below the lowest observed. \citet{aleman2014} only detected OH$^+$ emission from PNe with $T_* \ge 100 \kk$ - while our models in this temperature range are consistent with the observed surface brightness, we also predict similar values for $T_* = 80 \kk$, whereas observations by \citet{aleman2014} of three high-luminosity PNe with $80 \kk \le T_* < 100 \kk$ did not detect any OH$^+$ lines. The reasons for this are discussed in Section \ref{sec:80kk}. The CO $J=4-3$ surface brightnesses are essentially unchanged from the previous values. Figure \ref{fig:full617emis} shows the ArH$^+$ 617 GHz surface brightnesses versus $T_*$ for the models including the EUV flux. The $T_* = 150 \kk$ models both predict surface brightnesses above the detection threshold, while for lower $T_*$ the predicted values are much higher than the original case, although still consistent with the non-detection of this line in PNe. For HeH$^+$, the predicted $149 \um$ surface brightnesses are above the detection threshold for all models.

\begin{figure}
  \centering
  \includegraphics[width=\columnwidth]{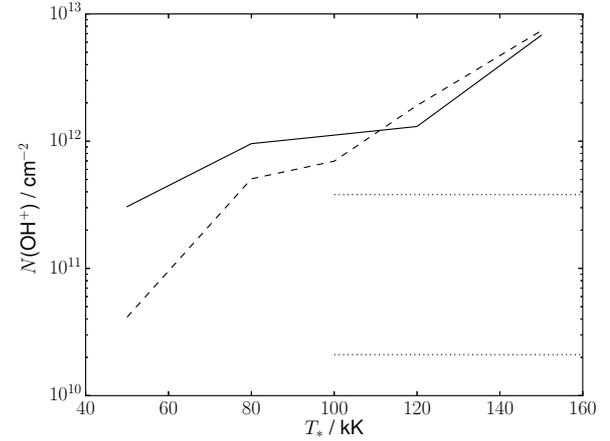}
  \caption{OH$^+$ column density versus $T_*$, for models with $\nh = 10^5 \pcc$, $\log g = 7.0$ and $L_* = 100 \lsun$ (solid line) and $1000 \lsun$ (dashed line) including the EUV flux. The dotted horizontal lines show the range of values calculated from observations by \citet{aleman2014}.}
  \label{fig:oh+abun}
\end{figure}

Figure \ref{fig:oh+abun} shows the OH$^+$ column density versus $T_*$ for models including the EUV flux. The column density is calculated by integrating through a single knot - multiple knots along the line of sight would result in higher values. The range of values in PNe calculated by \citet{aleman2014} from observations, assuming local thermodynamic equilibrium (LTE), are also shown. All models which reproduce the observed OH$^+$ line surface brightnesses have column densities significantly larger than those found by \citet{aleman2014}. \citet{otsuka2017} found an OH$^+$ column density of $10^{13} \pcs$ in their model of NGC 6781, also significantly higher than the value for this PN derived from observations. As we calculate the line emission directly, without assuming LTE, and since models without the EUV flux, which have lower OH$^+$ column densities, predict surface brightnesses below that observed, the higher values from our models are likely to be more accurate.

\section{Discussion}

\subsection{Location of the molecular emission}

\begin{figure}
  \centering
  \includegraphics[width=\columnwidth]{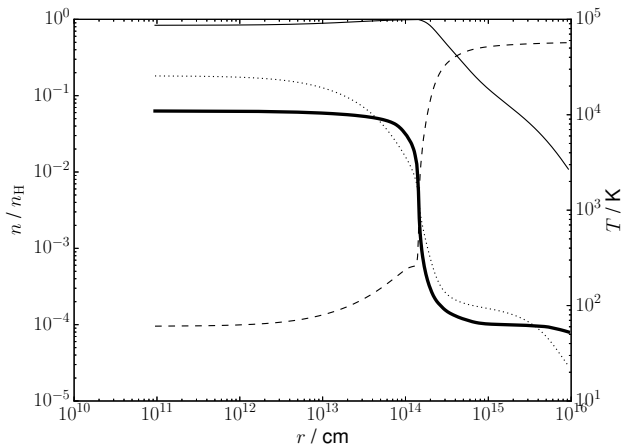}
  \caption{Abundances of H (thin solid line), H$_2$ (dashed line) and e$^-$ (dotted line) and gas temperature (thick solid line) versus distance into the knot for the $120 \kk$ $100 \lsun$ $\log g = 7.0$ EUV model. The left y-axis shows the abundance relative to hydrogen nuclei, and the right the gas temperature.}
  \label{fig:l2t120prop}
\end{figure}

Figure \ref{fig:l2t120prop} shows the abundances of H, H$_2$ and e$^-$, and the gas temperature, versus distance into the knot for the $120 \kk$ $100 \lsun$ {\sc ucl\_pdr} model including the EUV flux. At the edge of the knot the gas temperature and ionization fraction are high ($\sim 10^4$~K and $\sim 0.2$ respectively). The ionization fraction decreases smoothly with distance as the EUV flux is rapidly attenuated, while the gas temperature drops much more sharply at a distance of $\sim 2 \times 10^{14}$~cm. \citet{aleman2011}, who used similar values for the gas and central star properties, found similar temperatures and ionization fractions for their K1 model (with a step-function density profile), although the decrease in temperature is less sharp than in our models. The H$_2$ abundance rises slowly from an initial level of $10^{-4}$ before rapidly increasing at $\sim 2 \times 10^{14}$~cm, with the PDR region where hydrogen is mostly atomic taking up a relatively small part of the knot. \citet{aleman2011} stopped their models once the gas temperature reached $100 \kel$ - in this region, our models are in good agreement, although our H$_2$ abundances in the $100 \kel$ region are slightly lower.

\begin{figure}
  \centering
  \includegraphics[width=\columnwidth]{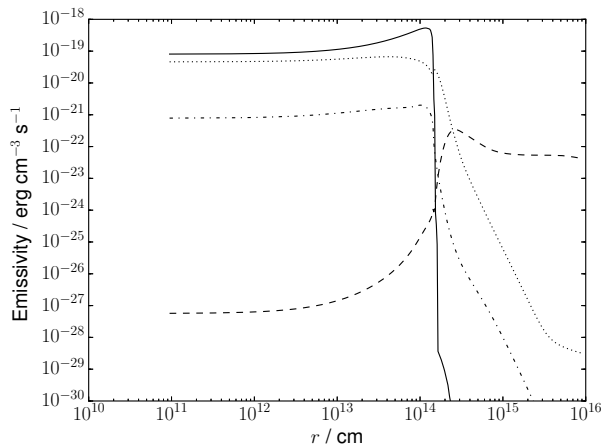}
  \caption{Line emissivities of H$_2$ $2.12 \um$ (solid line), CO $J=4-3$ (dashed line), OH$^+$ 971 GHz (dotted line) and ArH$^+$ 617 GHz (dot-dashed line) versus distance into the knot for the $120 \kk$ $100 \lsun$ $\log g = 7.0$ EUV model.}
  \label{fig:l2t120emis}
\end{figure}

Figure \ref{fig:l2t120emis} shows the emissivities of the H$_2$ $2.12 \um$, CO $J=4-3$, OH$^+$ 971 GHz and ArH$^+$ 617 GHz lines versus distance. The $2.12 \um$ emission is concentrated in the ionized and PDR regions, despite the H$_2$ abundance not rising above $10^{-3}$, and the emissivity is negligible in the molecular region of the knot as the much lower gas temperatures are unable to significantly populate the upper vibrational levels. \citet{aleman2011} found similar peak emissivities for this line, although in our models the emissivity drops off much more sharply into the cloud, as with the temperature. The OH$^+$ and ArH$^+$ emissivities are also highest at the edge of the knot, but decrease more gradually than for H$_2$, with the emission extending into the molecular region. The CO $J=4-3$ emissivity, unlike the other lines, is highest in the molecular region, with the contribution from the ionized/atomic parts of the knot negligible. Increasing $L_*$ produces higher temperatures and ionization fractions, and a smaller molecular region, while the size of the molecular region increases with $T_*$ as the FUV flux decreases.

\subsection{Treatment of EUV flux}
\label{sec:euv}

Our EUV models include, as part of the X-ray flux in {\sc ucl\_pdr}, the ionizing photon flux from $0.1-750$~\r{A} as X-rays in {\sc ucl\_pdr}. The $750-912$~\r{A} flux is absorbed in the ionized region treated by {\sc mocassin}. {\sc ucl\_pdr} uses the methods described by \citet{meijerink2005} to treat X-rays, which are not designed with EUV photons in mind. The code calculates the wavelength-dependent cross-section for photoionization using the fits from \citet{verner1995}, and uses this to calculate the ionization and heating rates at each point in the knot. As the fits are not limited to X-ray wavelengths, the contribution to the ionization and gas heating by the EUV radiation will be correctly included. However, the chemical network only includes singly- and doubly-ionized elements, and of these, only cooling by C$^+$, Si$^+$ and Fe$^+$ is accounted for. The cooling rate is therefore likely to be underestimated, and the model gas temperatures likely too high.

Molecular photodissociation rates are based on the UMIST database \citep{mcelroy2013} rates, scaled by the ratio of the energy density between $912$ and $2400$~\r{A} to that of the Draine field \citep{draine1978}, and so does not include the effects of EUV photons. Significantly for our results, OH$^+$ \citep{saxon1986}, ArH$^+$ \citep{roueff2014} and HeH$^+$ \citep{roberge1982} all have photodissociation cross-sections which peak at EUV wavelengths. The EUV flux is also not included in the treatments of dust heating and gas heating by the dust photoelectric (PE) effect, which may be important in the deeper regions of our models where the dust PE effect is the dominant heating mechanism.

EUV photons are attenuated rapidly in the knots - in both {\sc mocassin} and {\sc ucl\_pdr} the flux at these wavelengths is reduced to zero within $\sim 10^{-4} \pc$, so the effects are only significant in the outer regions of the knot (corresponding to the transition zones in \citet{aleman2011a}). As the CO emission originates from the PDR regions deeper into the knots, we do not expect those values to be sensitive to the EUV treatment - the differences between models with and without EUV are minimal. For the other molecules, the effects of including EUV are much more significant as the majority of the emission originates in the transition zones. In these parts of the knots, the heating rate is dominated by the X-ray/EUV flux, so the PE effect can be neglected. The main coolant in our models in these regions is atomic hydrogen - assuming that this would be the case even if cooling from all atomic/ionic species were included, we can be reasonably confident in the calculated temperatures in these regions.

The main issue with our treatment of the EUV flux is therefore the neglect of photodissociation by EUV photons, which will only have a significant impact if it is the main destruction mechanism for a particular molecule. \citet{aleman2004} and \citet{aleman2011} found charge exchange to be more significant in the destruction of H$_2$ in the transition zones of PNe, and as photodissociation by FUV photons, longwards of $912$~\r{A}, is accounted for, our H$_2$ abundances are unlikely to be strongly affected. For OH$^+$, ArH$^+$ and HeH$^+$, the increased photodissociation cross-sections in the EUV region (compared to the FUV) means that the photodissociation rates used are lower than the true values, and therefore the molecular abundances and the line emission will be overestimated. The extent will depend on the rates of other destruction mechanisms - in the transition zones, dissociative recombination with electrons is likely to be the most important destruction mechanism for the molecular ions, and if this is significantly higher than the photodissociation rates the effect on the abundances will be minimal.

\subsection{Effect of input parameters}

We have not investigated changing the values of a number of input parameters, including the dust-to-gas ratio, cosmic ray ionization rate and the distance of the knot from the ionizing source, $r_k$, from their 'standard' values, all of which may affect the resulting molecular emission. \citet{aleman2011} found that increasing the dust-to-gas mass ratio by an order of magnitude resulted in a $\sim 20$ per cent increase in H$_2$ $2.12 \um$ surface brightness in their models of cometary knots in NGC 7293. As this increase is most likely due to the increased H$_2$ formation rate on dust grains \citep{aleman2011a}, other molecules are unlikely to be as strongly affected, and it is not enough to remove the need for the EUV flux. Similarly, we find increasing the cosmic ray ionization rate by a factor of $100$ has little effect except for a moderate increase in the OH$^+$ surface brightness for the lowest $T_*$ models - with the inclusion of the EUV flux, the effects are negligible in all models.

\citet{aleman2011} also found that $r_k$ has an impact on the H$_2$ emission, which presumably applies to other molecules as well. The results shown in \citet{aleman2011} suggest that the H$_2$ surface brightness (for a given set of knot parameters) does not vary enormously with distance, unless the knot is beyond the ionization front, at which point it decreases rapidly. Our value of $r_k = 0.2 \pc$ is within the nebular ionization front radius for all central star parameters investigated. Reducing $r_k$ to $0.1 \pc$ for the $100 \lsun$ $120 \kk$ model, we find the H$_2$, OH$^+$, ArH$^+$ and HeH$^+$ surface brightnesses all increase by a factor of $\sim 2$, while the CO surface brightness decreases by a similar factor. \citet{aleman2011} also found that in some cases the H$_2$ surface brightness decreases with $r_k$ below a certain radius, presumably due to the increased UV flux dissociating the molecules. This behaviour might also be expected for OH$^+$, ArH$^+$ and HeH$^+$, although as discussed in Section \ref{sec:euv}, this is not treated in our models due to the non-inclusion of EUV photodissociation for these molecules.

\subsection{Shock heating and fluorescent emission}

With the inclusion of EUV photons, our models predict H$_2$ surface brightnesses for $T_* \ge 80 \kk$ within a factor of a few of the lowest values observed in PNe \citep{hora1999}. However, H$_2$ $2.12 \um$ emission is observed in PNe with a much wider range of central star temperatures, down to $48 \kk$ for NGC 40 \citep{hora1999,ueta2014}, and many PNe have surface brightnesses far higher than the values we obtain. This suggests that an additional source of H$_2$ emission is present in actual PNe which is missing from our models.

We assume that the PN knots are in thermal equilibrium, with the heating from the incident radiation field balanced by cooling from line emission and gas-grain interactions. H$_2$ emission can, however, be produced in the shocked region produced by the interaction of nebular material with the AGB wind from earlier evolutionary stages \citep{natta1998}, in which case the gas may be heated to higher temperatures. \citet{otsuka2017} proposed that the H$_2$ emission from NGC 6781 - one of the PNe with an OH$^+$ detection - is in fact due to shock heating, which they modelled by setting a minimum gas temperature. Table \ref{tab:helixmodels} lists observed molecular line surface brightnesses for NGC 6781 and NGC 7293, a PN with similar central star properties to those used by \citet{otsuka2017} ($100 \lsun$, $120 \kk$; \citealt{henry1999}) but with H$_2$ emission which does not appear to be caused by shocks \citep{odell2007}, along with predicted values from three PDR models with these central star parameters - our initial model (without any incident EUV radiation), the same model including the EUV flux and a model with both EUV flux and a minimum temperature $T_{\rm min} = 1500 \, {\rm K}$ (\citealt{otsuka2017} used $T_{\rm min} = 1420 \, {\rm K}$ for NGC 6781).

The initial model fails to reproduce the observed surface brightness of any line for both PNe, although the values for ArH$^+$ and HeH$^+$ are consistent with the non-detection of these lines. The EUV model is at worst within a factor of a few of observed surface brightnesses in NGC 7293 for H$_2$, CO and OH$^+$, but predicts much lower H$_2$ and CO surface brightnesses than observed in NGC 6781. \citet{aleman2011} also modelled the H$_2$ emission from the cometary knots in NGC 7293, and found surface brightnesses in good agreement with observed values, without incorporating shock heating, assuming the knot is seen edge on and integrating the peak emissivity over knot depths of order $\sim 0.001 \pc$. If we adopt this approach, rather than integrating the emissivity through the knot, we find a $2.12 \um$ surface brightness of $1.3 \times 10^{-5} \sfb$, within the range of observed values. The $T_{\rm min}$ model predicts H$_2$ and OH$^+$ surface brightnesses somewhat higher than observed in NGC 6871, while the CO $J=4-3$ line is $\sim 100$ times stronger than observed. The H$_2$ surface brightness is significantly higher than for NGC 7293 and for all of our other models, and is of the same order as the highest values observed in PNe \citep{hora1999}, suggesting that the emission from these object may originate from shocked gas. \citet{akras2017} found higher H$_2$ surface brightnesses (up to $10^{-3} \sfb$) in K $4-47$, which they suggest shows evidence of shock interactions. The ArH$^+$ and HeH$^+$ surface brightnesses are unchanged by the imposition of a minimum temperature. The ArH$^+$ 617 GHz surface brightness is below the detection threshold for both models, while the HeH$^+$ $149 \um$ surface brightness is higher.

\begin{table*}
  \centering
  \caption{Observed line surface brightnesses for NGC 7293 and NGC 6781, and models with $L_* = 100 \lsun$ and $T_* = 120 \kk$. Details of models are described in the text. Surface brightnesses are in $\sfb$. Observed values are from \citet{aleman2011} (H$_2$) and \citet{etxaluze2014} (CO, OH$^+$, ArH$^+$) for NGC 7293, and \citet{otsuka2017} (H$_2$, CO) and \citet{aleman2014} (OH$^+$, ArH$^+$, HeH$^+$) for NGC 6871.}
  \begin{tabular}{ccccccc}
    \hline
    Model & H$_2$ $2.12 \um$ & CO $J=4-3$ & OH$^+$ 971 GHz & ArH$^+$ 617 GHz & HeH$^+$ $149 \um$ \\
    \hline
    NGC 7293 & $0.9-3.7 \times 10^{-5}$ & $0.7-2.0 \times 10^{-7}$ & $4.1-9.3 \times 10^{-7}$ & $\lesssim 4 \times 10^{-8}$ & - \\
    NGC 6781 & $2.7 \times 10^{-4}$ & $3.7 \times 10^{-7}$ & $4.3-4.7 \times 10^{-7}$ & $\lesssim 5 \times 10^{-8}$ & $\lesssim 9 \times 10^{-7}$ \\
    Initial & $7.7 \times 10^{-17}$ & $3.4 \times 10^{-8}$ & $4.2 \times 10^{-10}$ & $2.5 \times 10^{-12}$ & $9.5 \times 10^{-12}$ \\
    EUV & $3.8 \times 10^{-6}$ & $4.2 \times 10^{-8}$ & $6.6 \times 10^{-7}$ & $1.8 \times 10^{-8}$ & $2.1 \times 10^{-6}$ \\
    $T_{\rm min}$ & $9.0 \times 10^{-4}$ & $2.3 \times 10^{-5}$ & $1.1 \times 10^{-6}$ & $1.8 \times 10^{-8}$ & $2.1 \times 10^{-6}$ \\
    \hline
  \end{tabular}
  \label{tab:helixmodels}
\end{table*}

An alternative excitation mechanism for H$_2$ line emission is fluorescence following the absorption of UV photons \citep{sternberg1989}, a mechanism which is not treated by {\sc ucl\_pdr}. \citet{odell2007} found that in the case of NGC 7293, fluorescence caused by non-ionizing UV photons (in the range $912-1100$~\r{A}) is an implausible mechanism for producing the observed H$_2$ emission, as too great a proportion of the flux must be reprocessed. For $\log g = 7.0$, the $T_* = 50 \kk$ model atmosphere emits the highest proportion of its flux in this range, about $0.14$, while for higher $T_*$ this decreases to $\lesssim 0.03$. The unattenuated flux at $0.2 \pc$ from a $100 \lsun$ central star is $0.08 \flux$, so the maximum possible energy available to power fluorescence for $T_* = 50 \kk$ is $0.01 \flux$. A $2.12 \um$ line surface brightness of $10^{-5} \sfb$ (the lowest observed values from \citet{hora1999}) would require $\sim 10^{-4} \flux$ if the emission is powered by fluorescence, corresponding to only $1 \%$ of the available flux, suggesting that H$_2$ emission from PN with $T_* \lesssim 100 \kk$ could plausibly originate from this mechanism. For higher $T_*$, the lower flux in the range producing fluorescence means that the criticisms of \citet{odell2007} still stand, and the highest H$_2$ surface brightnesses observed in PNe seem incapable of being produced by this mechanism.

\subsection{Non-detections of OH$^+$}
\label{sec:80kk}

In the sample of \citet{aleman2014}, all the PNe with OH$^+$ detections have central star temperature $T_* \gtrsim 100 \kk$. However, two PNe listed as having central star temperatures in this range, NGC 7027 and Mz 3, are not detected in OH$^+$. \citet{aleman2014} suggested that the high ($2.3$) C/O ratio in NGC 7027 may explain the non-detection of OH$^+$, as the available oxygen is locked up in CO molecules. Increasing the carbon abundance and reducing the oxygen abundance to $6 \times 10^{-4}$ and $2 \times 10^{-4}$ respectively, our $100 \lsun$, $120 \kk$ model predicts an OH$^+$ surface brightness of $2.4 \times 10^{-7} \sfb$, only slightly reduced from the original value of $6.6 \times 10^{-7} \sfb$. This reflects the lower oxygen abundance rather than any locking-up of the oxygen, as the OH$^+$ emission originates in the ionized/PDR region where the CO abundance is low. The lack of observed OH$^+$ emission in NGC 7027 is therefore unlikely to be due to the C/O ratio, and may instead be due to the smaller nebular radius ($0.03 \pc$ compared to $\sim 0.1 \pc$ for PNe with OH$^+$ detections), and its very high central star luminosity ($10^4 \lsun$; \citealt{middlemass1990}) which would imply a much higher flux of UV photons in the knot and therefore a higher photodissociation rate for OH$^+$. This could also account for the non-detection of ArH$^+$ and HeH$^+$ in this PN, which was in apparent conflict with the predictions of our models. Mz 3 was listed by \citet{aleman2014} as having $T_* = 107 \kk$, citing \citet{phillips2003}. However, a number of other authors have estimated a much lower central star temperature of $32-36 \kk$ \citep{cohen1978,zhang2002,smith2003}, for which our models predict no detectable OH$^+$ emission.

Our models with $T_* = 80 \kk$ predict OH$^+$ surface brightnesses similar to those with $T_* = 100 \kk$. However, of the three PNe listed by \citet{aleman2014} as having $T_*$ in the range of $80-90 \kk$, none have OH$^+$ detections. All three PNe (NGC 3242, NGC 7009 and NGC 7026) are highly luminous ($L_* \sim 2000-5000 \lsun$; \citealt{frew2008}), and have ionized radii of $\sim 0.1 \pc$, smaller than assumed in our low luminosity models, so the increase in the UV flux, and therefore the photodissociation rates, will be very substantial, accounting for their lack of detectable molecular line emission.

\subsection{ArH$^+$ and HeH$^+$ emission}

Our EUV models all predict HeH$^+$ surface brightnesses at or above detection thresholds, and our models with $T_* = 150 \kk$ also predict detectable levels of ArH$^+$ emission, despite neither molecule having yet been detected in PNe. HeH$^+$ has previously been predicted to form in detectable quantities in PNe (e.g. \citealt{cecchi1993}), while ArH$^+$ is known to be present in mainly atomic regions with low H$_2$ abundances and a suitable ionization source \citep{schilke2014,priestley2017}, conditions which certainly apply to the surface regions of knots in PNe.

\begin{figure}
  \centering
  \includegraphics[width=\columnwidth]{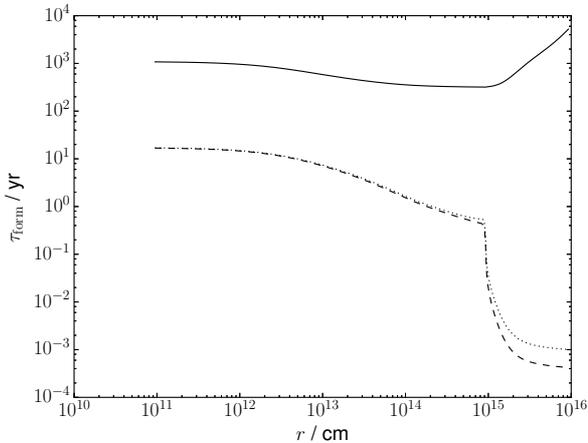}
  \caption{Formation timescales versus distance into the knot for HeH$^+$ (solid line), OH$^+$ (dashed line) and ArH$^+$ (dotted line), for the $150 \kk$ $1000 \lsun$ $\log g = 7.0$ EUV model.}
  \label{fig:timescales}
\end{figure}

In models for the clumps in the Crab Nebula \citep{priestley2017}, it was found that the predicted equilibrium level of HeH$^+$ emission was large enough to have been detected in Herschel PACS spectra. However, the formation timescale for HeH$^+$ was found to be significantly larger than the age of the Crab, suggesting that the assumption of chemical equilibrium can lead to an overestimate for the molecule's abundance. The formation timescacle for ArH$^+$, by contrast, is short enough that the assumption of equilibrium is justified. Figure \ref{fig:timescales} shows the formation timescales for HeH$^+$, OH$^+$ and ArH$^+$ versus distance into a knot at a distance of $0.2 \pc$, for the $150 \kk$, $1000 \lsun$ EUV model. The abundances of all three molecules drop off beyond $\sim 10^{15} \, {\rm cm}$, which combined with the lower temperature means that there is no significant emission beyond this point. The formation timescales in this region for ArH$^+$ and OH$^+$ are similar, of order a few to ten years, whereas for HeH$^+$ the timescale is much larger ($\tau_{\rm form} \sim 300-1000 \, {\rm yr}$). Given typical ages of these old PNe (>1000 yr; \citet{ueta2014}), this is unlikely to be long enough to significantly reduce the abundance of these species. A possible explanation is the EUV field - as discussed in Section \ref{sec:euv}, photodissociation by EUV photons is not treated by our models. The inclusion of the EUV flux would therefore be expected to reduce the abundances of both molecules, potentially reducing the emission to below the detection threshold.

\section{Conclusions}

We have performed combined photoionization and PDR modelling of PN cometary knots, for a range of central star parameters and gas densities. Our initial PDR models fail to reproduce the observed H$_2$ and OH$^+$ surface brightnesses for PNe with $T_* < 150 \kk$. Only by including the ionizing UV flux from the central star do the models predict surface brightnesses for $T_* \ge 100 \kk$ consistent with the values observed in PNe. Predicted OH$^+$ column densities are $\gtrsim 10^{12} \pcs$, significantly larger than those derived from observations assuming LTE, but in agreement with previous modelling work \citep{otsuka2017}. Predicted H$_2$ surface brightnesses are somewhat lower than observed values, and our models do not explain the highest emission strengths observed in PNe, or the detection of H$_2$ emission in PNe with cooler ($T_* < 80 \kk$) central stars. The presence of shocks may be capable of producing the higher H$_2$ surface brightnesses observed, while fluorescence due to excitation by UV photons could be a viable source of H$_2$ emission for nebulae with central star temperatures below $50 \kk$. Our models predict HeH$^+$ (and, for $T_* = 150 \kk$, ArH$^+$) line surface brightnesses potentially above detection thresholds, despite neither molecule having been detected in PNe. As photodissociation by EUV photons is not treated for either molecule, the true surface brightnesses may be significantly lower than our predicted values if this is an important destruction mechanism.

\section*{Acknowledgements}

FP is supported by the Perren fund and UCL IMPACT fund. MJB acknowledges support from the European Research Council grant SNDUST ERC-2015-AdG-694520. We would like to thank Prof. Jonathan Tennyson for providing the recent ArH$^+$ dissociative recombination rates of Abdoulanziz et al. The TheoSSA service (http://dc.g-vo.org/theossa) used to retrieve theoretical spectra for this paper was constructed as part of the activities of the German Astrophysical Virtual Observatory.




\bibliographystyle{mnras}
\bibliography{pn}





\bsp	
\label{lastpage}
\end{document}